\documentclass[prx,twocolumn,citeautoscript,amsmath,superscriptaddress]{revtex4-1}

\usepackage{graphicx}
\usepackage{newtxtext}
\usepackage{epsfig}
\usepackage{amsmath,bbm,amssymb,bm,times}

\usepackage[colorlinks,linkcolor=blue,citecolor=blue]{hyperref}

\begin{document}
\title{Exceptional mode topological surface laser}
\author{Kazuki Sone}
\email{sone@noneq.t.u-tokyo.ac.jp}
\affiliation{Department of Applied Physics, The University of Tokyo, 7-3-1 Hongo, Bunkyo-ku, Tokyo 113-8656, Japan}
\author{Yuto Ashida}
\affiliation{Department of Physics, The University of Tokyo, 7-3-1 Hongo, Bunkyo-ku, Tokyo 113-0033, Japan}
\affiliation{Institute for Physics of Intelligence, The University of Tokyo, 7-3-1 Hongo, Tokyo 113-0033, Japan}
\author{Takahiro Sagawa}
\affiliation{Department of Applied Physics, The University of Tokyo, 7-3-1 Hongo, Bunkyo-ku, Tokyo 113-8656, Japan}
\affiliation{Quantum-Phase Electronics Center (QPEC), The University of Tokyo, 7-3-1 Hongo, Bunkyo-ku, Tokyo 113-8656, Japan}

\begin{abstract} 
Band topology has been studied as a design principle of realizing robust boundary modes. Here, by exploring non-Hermitian topology, we propose a three-dimensional topological laser that amplifies surface modes. The topological surface laser is protected by nontrivial topology around branchpoint singularities known as exceptional points. In contrast to two-dimensional topological lasers, the proposed three-dimensional setup  can realize topological boundary modes without judicious gain at the edge or symmetry protection, which are thus robust against a broad range of disorders. We also propose a possible optical setup to experimentally realize the topological surface laser. Our results provide a general guiding principle to construct non-Hermitian topological devices in three-dimensional systems.
\end{abstract}

\maketitle

\section{Introduction\label{section1}}
Topology of band structures has attracted much interest in condensed matter physics owing to its robust properties exemplified by the associated boundary modes \cite{Kane2005,Bernevig2006,Hasan2010,Qi2011}. Recent studies have extended the notion of topology to non-Hermitian Hamiltonians \cite{Martinez2018,Xiong2018,Shen2018,Kunst2018,Yao2018,Gong2018,Yokomizo2019,Zhou2019,Kawabata2019a,Okuma2020,Zhang2020}, which can describe dissipative systems. Non-Hermitian systems can exhibit novel topological states characterized by the winding number of eigenvalues \cite{Gong2018}. Such winding number plays crucial roles in, e.g., localization of bulk modes called skin effect \cite{Martinez2018,Xiong2018,Yao2018,Kunst2018,Yokomizo2019,Okuma2020,Zhang2020} and the protection of exceptional points (EPs) \cite{Kato1966,Dembowski2001,Xu2017,Zhou2018,Okugawa2019,Budich2019,Kawabata2019b} where two or more eigenvectors coalesce. 

From a broad perspective, non-Hermitian topology influences a variety of fields in physics, including photonics \cite{Weimann2017,Zhou2018,Ota2020}, electrical circuits \cite{Ezawa2019,Helbig2020}, quantum walks \cite{Mochizuki2016,Weidemann2020,Xiao2020}, and biological systems \cite{Murugan2017,Yamauchi2020,Tang2021,Yoshida2021}. One of the promising applications of non-Hermitian topology is a topological laser \cite{St-Jean2017,Harari2018,Bandres2018,Song2020,Sone2020} which amplifies the boundary modes. One can construct such topological laser by introducing judicious gain at the edge of the sample \cite{Harari2018,Bandres2018}. Another route to realize a topological laser is edge modes protected by EPs in their dispersion relations \cite{Sone2020}. Such edge modes are termed as exceptional edge modes because they are robust even with topologically trivial bulk due to the topological protection of EPs at the edge.

While the pioneering work of band topology analyzes the quantum Hall effect in two-dimensional semiconductors \cite{Klitzing1980,Thouless1982}, many studies have also focused on three-dimensional topological materials \cite{Fu2007,Chen2009,Ringel2012,Slobozhanyuk2017,Lu2018,Yang2019}. Three-dimensional Hermitian systems can exhibit unique topological features without low-dimensional counterparts (cf.~Weyl semimetals \cite{Wan2011,Yang2011,Armitage2018}), which expand the application of topological materials. In contrast, despite their broadness and importance, topology of three-dimensional non-Hermitian systems \cite{Yang2019b,Carlstrom2019,Denner2021,Ghorashi2021} is less understood than Hermitian ones. Especially, topological lasers in three-dimensional systems are so far unexplored, while such extension is crucial to design a large-scale topological laser. 

In this paper, we propose a topological laser that exhibits lasing wave packets localized at the surface of a three-dimensional system. In contrast to the two-dimensional counterparts investigated in previous research \cite{Harari2018,Bandres2018,Kawabata2019a,Song2020,Sone2020}, the proposed topological laser requires neither judicious gain nor symmetries and thus is robust against symmetry-breaking and edge-distorting disorders. One can realize such topological laser by utilizing exceptional surface modes protected by EPs in the surface band. We construct a prototypical model of the topological surface laser by introducing non-Hermitian coupling to a weak topological insulator \cite{Fu2007,Ringel2012}. By numerically calculating a band structure and linear dynamics, we confirm the existence of lasing surface modes (surface-localized eigenstates that are amplified and accompany stationary or suppressed bulk modes \cite{Kawabata2019a,Song2020,Sone2020}) protected by EPs. The topological surface laser can be realized in an optical setup utilizing stacked ring-resonator arrays. Our results reveal the potential of three-dimensional non-Hermitian topology toward realizing topological surface laser.

The rest of the paper is organized as follows. In Sec.~\ref{section2}, we analyze prototypical models of the topological surface laser. Then, we theoretically propose a photonic setup in Sec.~\ref{section3}. We also numerically demonstrate the existence of amplified surface modes in the effective Hamiltonian of the proposed setup. In Sec.~\ref{section4}, we discuss the no-go theorem of a single pair of EPs under the fully open boundary conditions, which indicates that the four EPs emerging in our models are minimum to protect the surface modes.  Section \ref{section5} summarizes the results and discusses the difference from previous studies. Some details are discussed in Appendix.

\begin{figure}[t]
  \includegraphics[width=70mm,bb=0 0 510 615,clip]{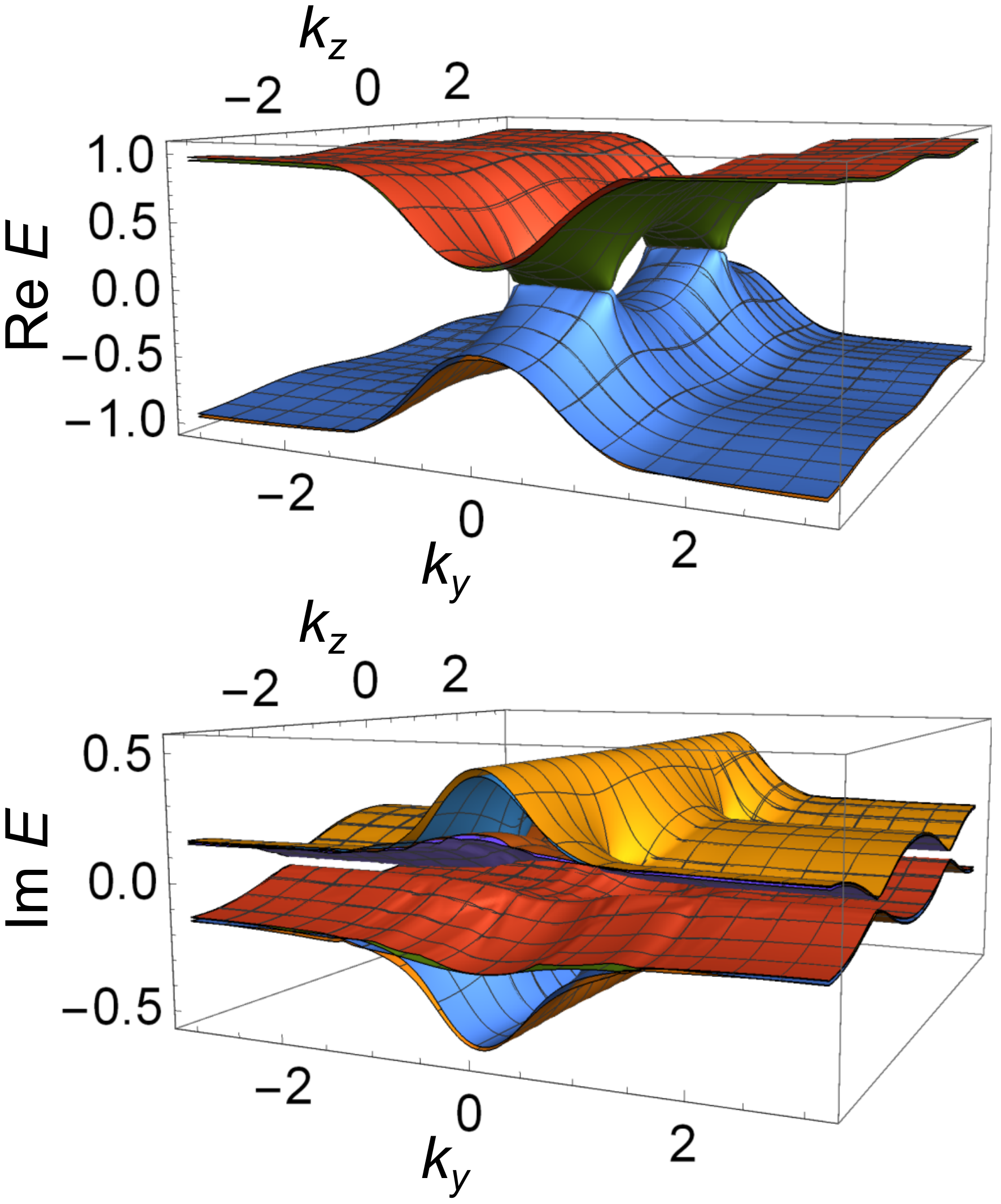}
  \caption{\label{fig1} Surface band structure of exceptional surface modes. The upper (lower) figure shows the real (imaginary) parts of complex dispersion relations of our model \eqref{prototypical_model}. For clarity, we only present the bands up to the fourth closest to the ${\rm Re}\,E=0$ plane in the upper panel. In the lower panel, we show eight bands with up to the fourth largest and smallest imaginary parts of eigenvalues. The parameters used are $u=-1$, $c=0.3$, $\gamma=0.5$, $W_u=0.1$, $W_i=0.1$, and $W_a=0.1$.
} 
\end{figure}

\section{Model analyses\label{section2}}
\subsection{Prototypical model\label{sec2a}}
In general, one can construct a three-dimensional topological surface laser by introducing a non-Hermitian spin coupling to a weak topological insulator. Here, we consider a weak topological insulator composed of stacked layers of spin Hall systems. Such a weak topological insulator exhibits surface modes with two Dirac cones on its particular sides \cite{Fu2007}. Then, the non-Hermitian spin coupling splits those Dirac cones into pairs of EPs \cite{Zhou2018,Bergholtz2021} and nonzero imaginary parts of eigenvalues around them, which leads to topological protection and amplification of the surface modes.

By following the above procedure, we construct a prototypical model of topological surface lasers and numerically analyze its surface modes. We first consider the Chern insulator model $H_{\rm Chern} = (u+\cos k_x + \cos k_y) \sigma_z + \sin k_x \sigma_x + \sin k_y \sigma_y$ \cite{Qi2006} corresponding to a lattice version of the Dirac Hamiltonian, which describes low-energy dispersions of topological insulators \cite{Hasan2010,Qi2011}. We combine the Chern insulator and its time-reversal counterpart and obtain a quantum spin Hall system \cite{Bernevig2006}. Next, by stacking layers of such quantum spin Hall systems in the $z$ direction, we construct a weak topological insulator \cite{Fu2007,Ringel2012}. Finally, by introducing non-Hermitian coupling between time-reversal counterparts we obtain the following toy model described in the wavenumber space as
\begin{eqnarray}
 H(\mathbf{k}) &=& (u+\cos k_x + \cos k_y) I \otimes \sigma_z + \sin k_x \sigma_z \otimes \sigma_x \nonumber\\
 &{}& + \sin k_y I \otimes \sigma_y + (c\cos k_z+i\gamma) \sigma_y \otimes \sigma_x, \label{prototypical_model}
\end{eqnarray}
where $\sigma_i$ represents the Pauri matrices and $I$ is the $2\times2$ identity matrix. All the parameters $u$, $c$, $\gamma$ are real. $u+2$ determines the strength of on-site potentials corresponding to effective mass around $\mathbf{k}=\mathbf{0}$. $i\gamma \sigma_y \otimes \sigma_x$ represents the non-Hermitian coupling. The term proportional to $c\cos k_z$ is the Hermitian interlayer coupling, since it represents hoppings between neighbor sites in the $z$ direction.

Numerically calculating the surface band structure, we check the emergence of surface modes exhibiting lasing behavior in our model \eqref{prototypical_model}. We set open boundaries along the $y$-$z$ plane and impose the periodic boundary conditions in the other directions. In such boundary conditions, the wavenumbers parallel to the $y$ and $z$ directions, $k_y$ and $k_z$, are still good quantum numbers. Therefore, we can plot the dispersion relations as functions of $k_y$ and $k_z$, which we here call surface band structure. We also introduce a complex on-site random potential and a random spin coupling to investigate the robustness of gapless surface modes. Specifically, we add diagonal terms whose intensities are determined so that their real and imaginary parts follow uniform distributions ranging $[-W_{u,i},W_{u,i}]$ respectively and off-diagonal terms that couple the first and fourth components or the second and third ones with the random intensities following uniform distributions ranging $[-W_{a},W_{a}]$. We use the parameters $u=-1$, $c=0.3$, $\gamma=0.5$, $W_u=0.1$, $W_i=0.1$, and $W_a=0.1$ to clearly show the appearance of the surface modes protected by EPs, while the qualitative behaviors of surface modes are independent of the choice of the parameters unless it changes the topology of the system. 

Figure \ref{fig1} shows the obtained surface band structure. We confirm the existence of gapless surface modes with four EPs. Since some of those surface modes exhibit larger imaginary parts of eigenvalues than those of the bulk modes, they are amplified faster than the bulk modes and thus lead to lasing of surface sites. These gapless modes are stable against the disorders due to the topological protection of EPs in the surface bands \footnote{Disorders similar to those considered here can break symmetries essential to symmetry-protected topological lasers and thus make two-dimensional topological lasers unstable \cite{Sone2020}.}. In this respect, the three-dimensional topological laser is more advantageous than its two-dimensional counterparts, because its protection requires no symmetries, and such symmetry-independent topological protection is possible only in the surface bands of three-dimensional systems which can have nonzero winding numbers around EPs \cite{Kawabata2019b}.

\begin{figure}[t]
  \includegraphics[width=70mm,bb=0 0 510 620,clip]{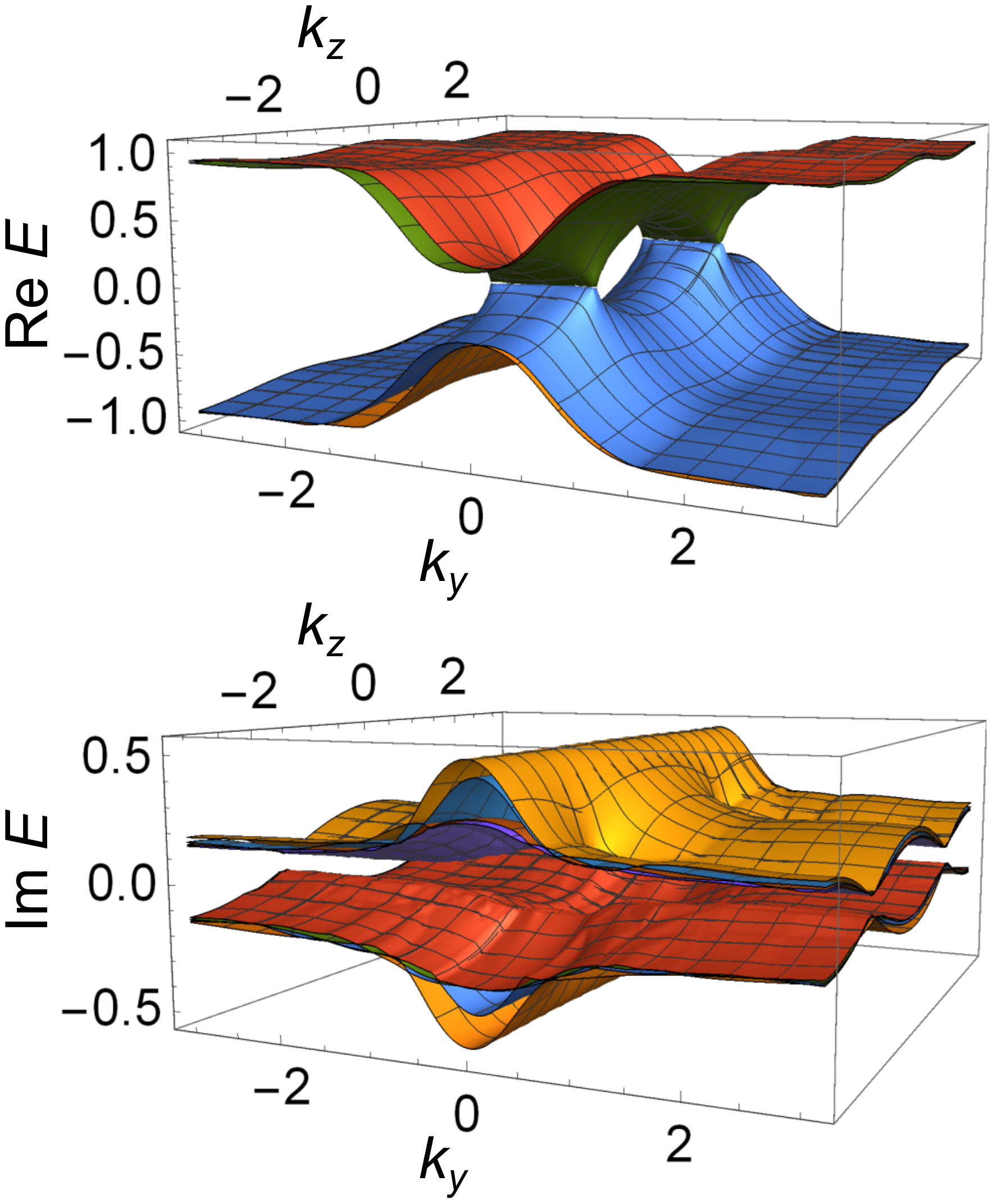}
\caption{\label{fig2} Band structure of the Hamiltonian of our model under the existence of all possible intracell disorders. We impose the open boundary condition in the $x$ direction. The real (imaginary) parts of eigenvalues are shown in the upper (lower) panel. The gapless modes protected by EPs robustly exist against the disorders. The parameters used are $u=-1$, $c=0.3$, $\gamma=0.5$, and $W=0.1$.
}
\end{figure}

It is noteworthy that since we construct the topological surface laser based on a weak topological insulator, gapless modes disappear in the surfaces parallel to the stacked layers \cite{Fu2007,Ringel2012}. One can confirm such anisotropy by numerical calculations of the surface band structure under the open boundary condition in the $z$ direction (cf.~Appendix \ref{appx_B}). We also note that it is essential to introduce the non-Hermitian coupling that has a similar form to the Hermitian interlayer coupling as in our prototypical model \eqref{prototypical_model}. On the one hand, if we introduce a different type of non-Hermitian coupling, we can realize the exceptional surface modes with exceptional rings or nodal lines \cite{Xu2017,Okugawa2019,Budich2019} (cf.~Appendix \ref{appx_C}). On the other hand, however, such exceptional surface modes are fragile against symmetry-breaking disorder as is the case with exceptional edge modes in two-dimensional systems \cite{Sone2020}. In contrast, the lasing surface modes in Fig.~\ref{fig1} are robust against such disorder, which is of significant advantage to construct topological lasers in practice.

We can further confirm the robustness of the lasing surface modes against a broad range of disorders by calculating surface band structure under the existence of all possible noises in the intracell interactions and on-site potentials. Since our model \eqref{prototypical_model} has four sublattices, the noises can be described by a random $4\times4$ matrix at each lattice point. We determine each component so that its real and imaginary parts follow the uniform distribution ranging $[-W,W]$ respectively. These noises include the noises considered in Fig.~\ref{fig1} (random on-site potentials and a random spin coupling) and can break any symmetries that play important roles in the classification of non-Hermitian topology \cite{Zhou2019,Kawabata2019a}. However, since the proposed topological surface lasers necessitate no symmetries, they are robust against these intracell noises.

We calculate the band structure of the surface modes under the existence of the noises. Here we assume that the noises are periodic in the $y$ and $z$ directions so that we can define the band structure. Figure \ref{fig2} shows the obtained band structure. We still find gapless surface modes with EPs, which indicates the robustness of the surface modes against all possible noises in the intracell interactions and on-site potentials.

\begin{figure}[b]
  \includegraphics[width=70mm,bb=0 0 505 595,clip]{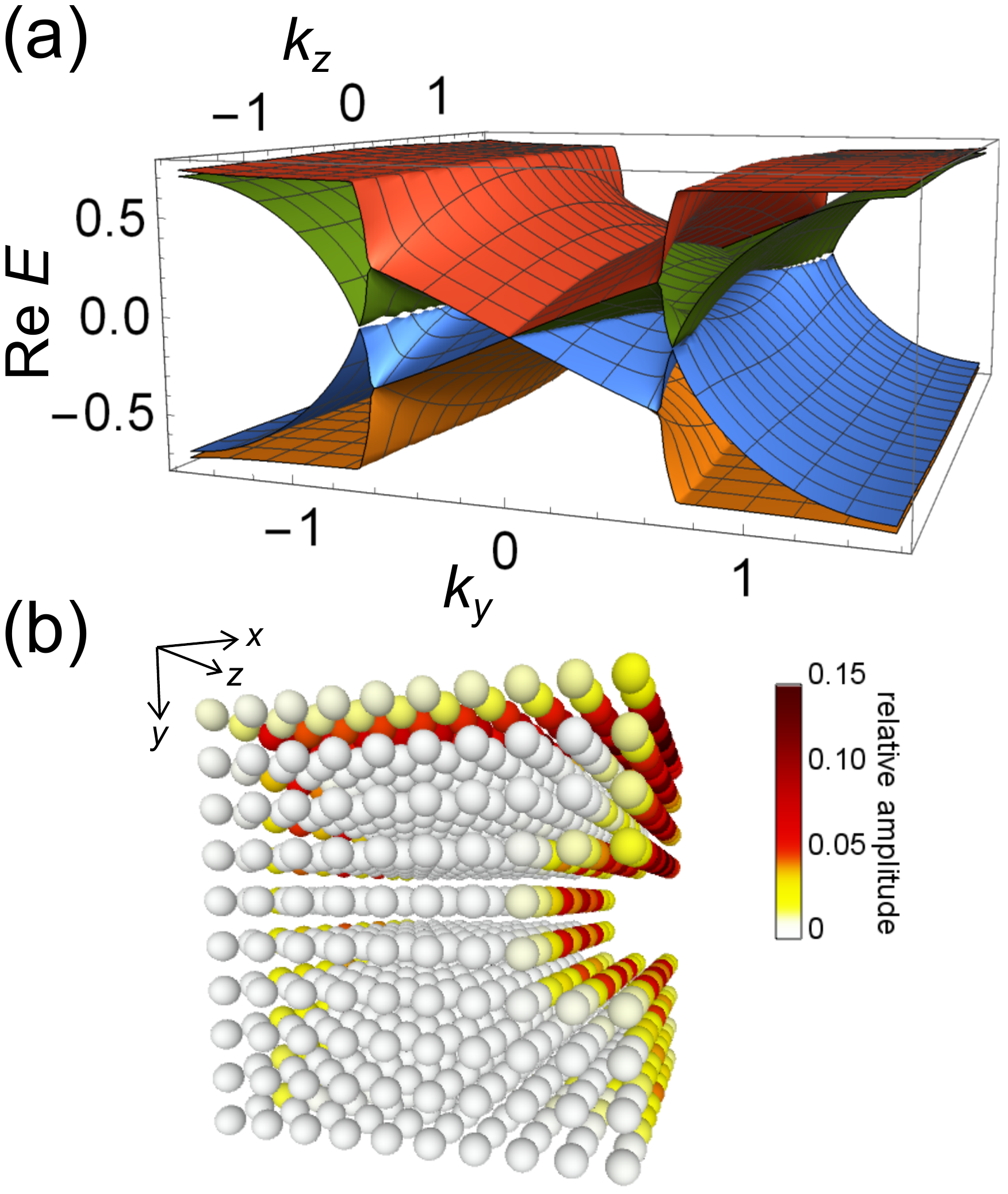}
  \caption{\label{fig3} Propagating lasing surface modes. (a) Surface band structure of exceptional surface modes with nonzero group velocity. The parameters used are $u=-1$, $a=1.5$, $b=0.5$, $c=0.2$, and $\gamma=0.9$. (b) Dynamical demonstration of propagating lasing surface modes. The color of each sphere represents the relative amplitude at each site, which is normalized so that the sum of squares is unity. We use the same parameters as panel (a).}
\end{figure}

\subsection{Propagating lasing surface modes\label{sec2b}}
Nonzero group velocities of lasing surface modes enable the robust transfer of lasing wave packets along the surface of a topological laser \cite{Harari2018,Bandres2018,Sone2020}. One can construct exceptional surface modes with nonzero group velocities by introducing asymmetry between effective spin degrees of freedom. By introducing such asymmetry, we can realize the difference between the slopes of the dispersions of the surface modes in each spin sector. Then, when the non-Hermitian interaction couples those surface modes, nonzero group velocities remain and are equal to the difference in the slopes of the original linear dispersions (similar discussions can be seen in \cite{Sone2020}). To demonstrate the possibility of the propagating lasing surface modes, we analyze the following Hamiltonian,
\begin{eqnarray}
 H(\mathbf{k}) &=& \left(
  \begin{array}{cc}
   (a+b) H_{\rm Chern} & (-ic\cos k_z+\gamma) \sigma_x \\
   (ic\cos k_z-\gamma) \sigma_x & (a-b) H_{\rm Chern}^{\ast} 
  \end{array}
  \right) \nonumber\\
 &=& (u+\cos k_x + \cos k_y) \left( a I \otimes \sigma_z + b \sigma_z \otimes \sigma_z \right) \nonumber\\
 &{}& + \sin k_x \left( b I \otimes \sigma_x + a \sigma_z \otimes \sigma_x \right) \nonumber\\
 &{}& + \sin k_y \left( a I \otimes \sigma_y + b \sigma_z \otimes \sigma_y \right) \nonumber\\
 &{}& + (c\cos k_z+i\gamma) \sigma_y \otimes \sigma_x. \label{propagating_surface_modes}
\end{eqnarray}
We construct this Hamiltonian by introducing different hopping amplitudes between the (effective) spin degrees of freedom, which are represented as $a+b$ and $a-b$. By numerically calculating the surface band structure of the pure system (the numerical method is the same as in Fig.~\ref{fig1}), we confirm the emergence of exceptional surface modes with nonzero group velocity $\partial {\rm Re}\,E/\partial k_y\neq 0$ (see Fig.~\ref{fig3}a).

We also demonstrate the unidirectional propagation of the lasing wave packets by simulating the dynamics of the Hamiltonian \eqref{propagating_surface_modes}. Here, we obtain the real-space description of the Hamiltonian via the inverse Fourier transformation and perform the fourth-order Runge-Kutta simulation, setting the time step as $dt=0.001$. Figure \ref{fig3}b shows a snapshot of the relative-amplitude distribution. We consider a $10\times10\times10$ cubic lattice with open boundaries and deform the boundary configuration, which is represented as the empty sites in the figure. As the initial condition, we excite the surface sites whose $x$ and $y$ components are $x=5$ and $y=10$. One can confirm that an amplified wave packet propagates along the surface parallel to the $z$ axis.

\begin{figure}[t]
  \includegraphics[width=60mm,bb=0 0 450 340,clip]{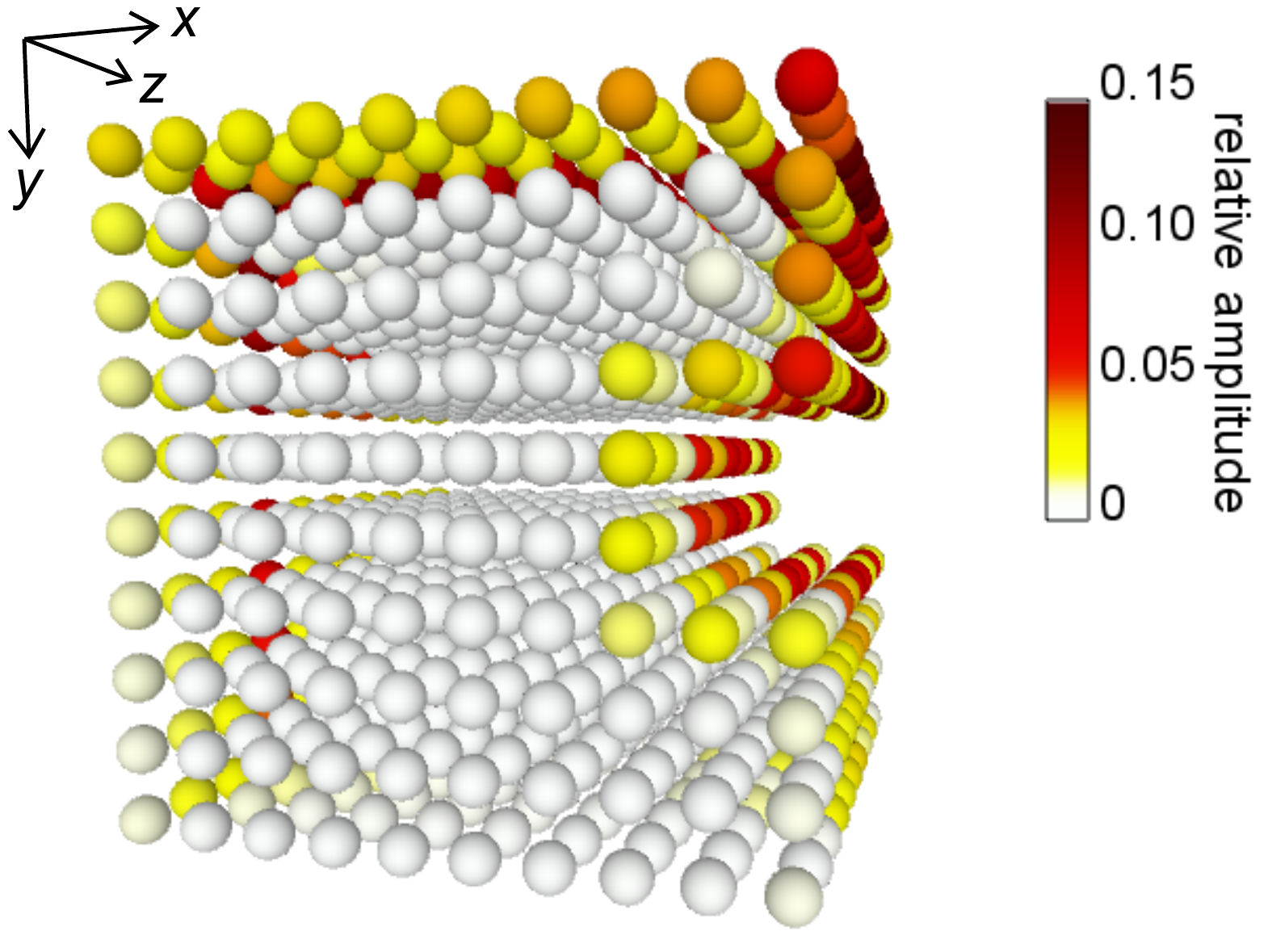}
\caption{\label{fig4} Propagating lasing surface modes under the existence of all possible intracell disorders. The color of each sphere represents the relative amplitude at each site, which is normalized so that the sum of squares is unity. One can observe the propagation of the surface-localized wave packet even under the existence of the intracell disorders. The parameters used are $u=-1$, $a=1.5$, $b=0.5$, $c=0.2$, $\gamma=0.9$, and $W=0.1$.
}
\end{figure}

We note that the lasing wave packet is transferred without backscattering even at the defect in the surface configuration. Thus, the lasing surface modes are robust against disorder on the surface of the system. We can also confirm the robustness against spatial modulation of the parameters as in the surface band structures as is discussed in the following paragraph. Furthermore, the robust surface modes under the fully open boundary conditions imply that they are immune to the non-Hermitian skin effect \cite{Martinez2018,Xiong2018,Yao2018,Kunst2018,Yokomizo2019,Okuma2020,Zhang2020}, which can affect other non-Hermitian surface modes (cf.~Appendix \ref{appx_C}).

We also calculate the dynamics of the lasing wave packets under the existence of all possible noises in the intracell interactions and on-site potentials as in Fig.~\ref{fig2}. We consider the Hamiltonian \eqref{propagating_surface_modes} with disorder terms. Figure \ref{fig4} shows the snapshot of the relative-amplitude distribution obtained by the simulation. One can confirm that an amplified wave packet propagates along the surface of the system without backscattering. The system has no $\mathbb{Z}_2$ or translation symmetries, while the obtained robust surface wave packets imply that the protection of the proposed surface modes is irrelevant to those symmetries.

\begin{figure}[b]
  \includegraphics[width=86mm,bb=0 0 1555 1650,clip]{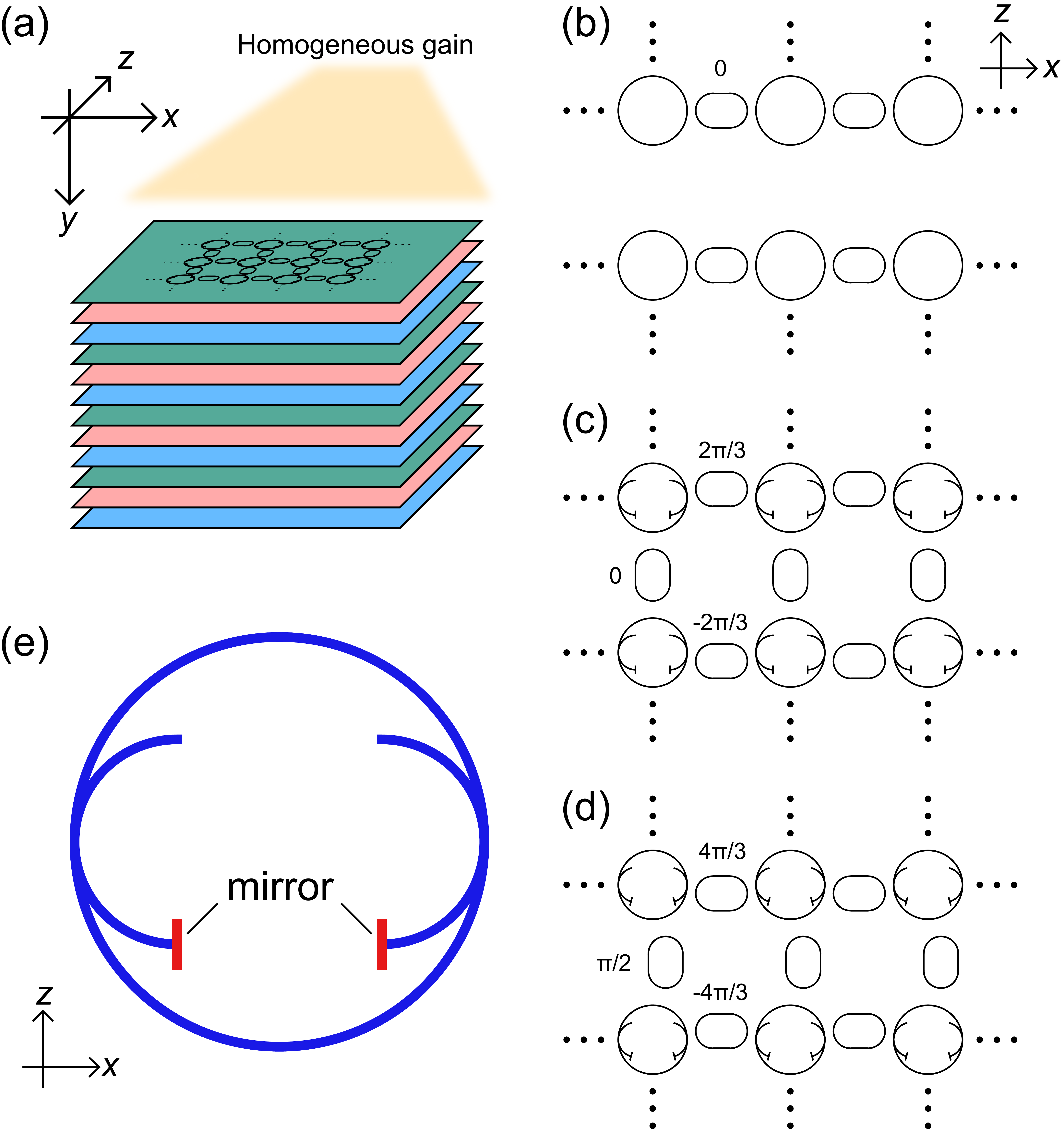}
  \caption{\label{fig5} Proposed optical setup of the topological surface laser. (a) Overview of the proposed ring-resonator arrays. Each color of a layer represents the distinct alignment of ring resonators which corresponds to one of those in panels (b)-(d). (b)-(d) Alignments of ring resonators in the layers of the proposed optical system. The numbers show the phase obtained via the wave propagation of clockwise modes in each waveguide. Counterclockwise modes acquire the opposite amount of phases via wave propagation. (e) Ring resonator to realize the non-Hermitian pseudo-spin coupling seen in panels (c) and (d). Each attached waveguide has a mirror at either end and realizes the non-Hermitian coupling.} 
\end{figure}

\section{Optical setup of a topological surface laser\label{section3}}
We next propose a possible optical implementation of a topological surface laser utilizing exceptional surface modes. We construct the topological surface laser by considering a photonic counterpart of a weak topological insulator and introducing a non-Hermitian coupling. A ring-resonator array is a typical setup to construct a counterpart of a quantum spin Hall system \cite{Hafezi2011,Hafezi2013,Ozawa2019}. In a ring resonator, clockwise and counterclockwise modes play effective spin degrees of freedom. Piling the ring-resonator arrays imitating quantum spin Hall systems, one can realize a weak topological system. Furthermore, a non-Hermitian spin coupling can also be introduced by using waveguides with mirrors and homogeneous gain.

\begin{figure}[t]
  \includegraphics[width=60mm,bb=0 0 510 710,clip]{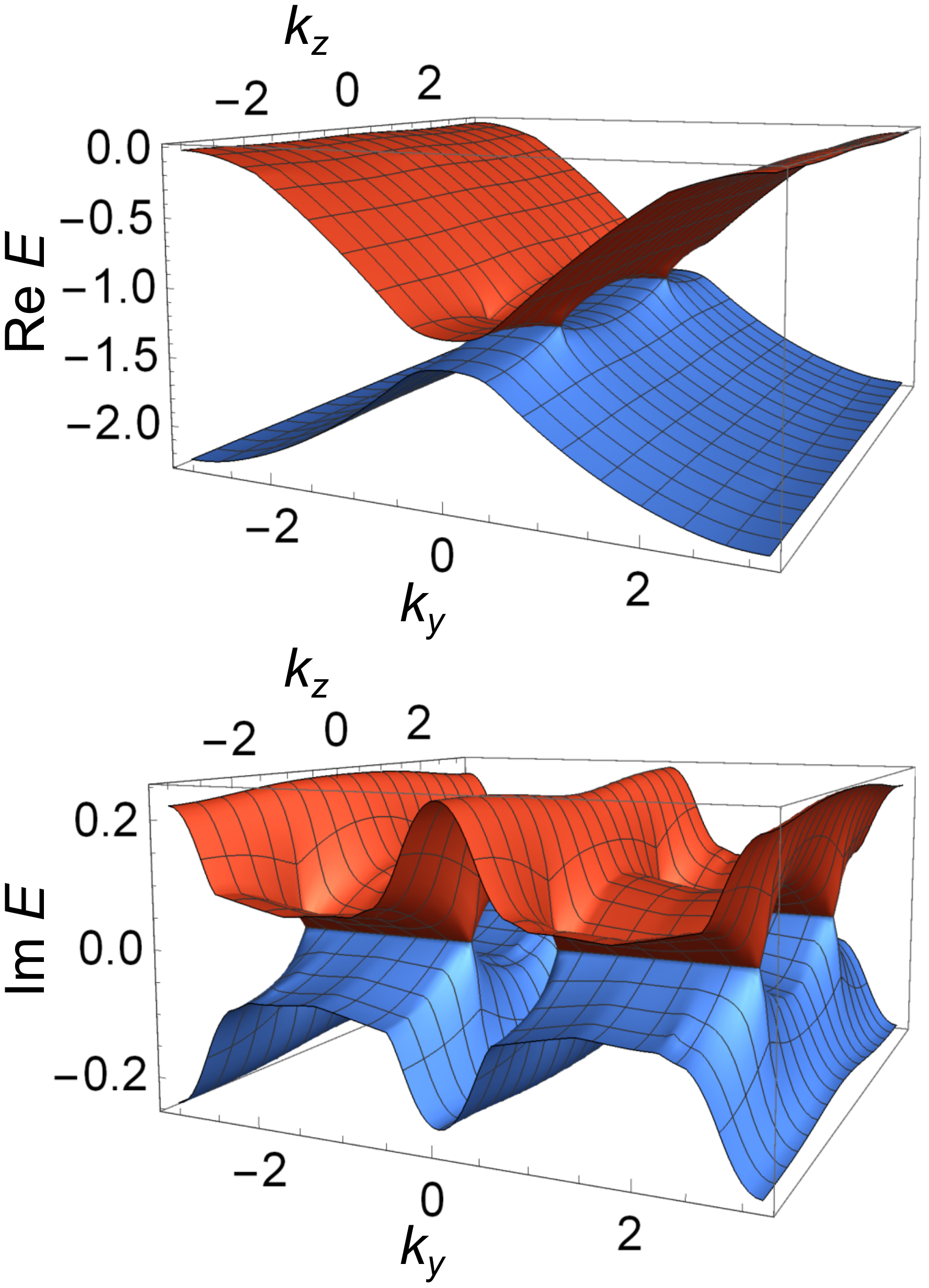}
\caption{\label{fig6} Surface band structure of the lasing surface modes in the effective Hamiltonian of the proposed photonic system. The upper (lower) panel shows the real (imaginary) parts of the surface bands. For clarity, we only show the bands up to the fourth closest to the bulk band gap around ${\rm Re}\,E=-1$ in the upper panel and four eigenvalues with up to the second largest and smallest imaginary parts in the lower panel. One can confirm the existence of gapless modes with four EPs in their surface bands. The gapless bands also exhibit larger imaginary parts than the bulk ones, which leads to lasing of surface sites. The parameters used are $a_x=a_y=1$, $a_z=0.1$, and $\beta=0.5$.
}
\end{figure}

\begin{figure}[t]
  \includegraphics[width=70mm,bb=0 0 660 640,clip]{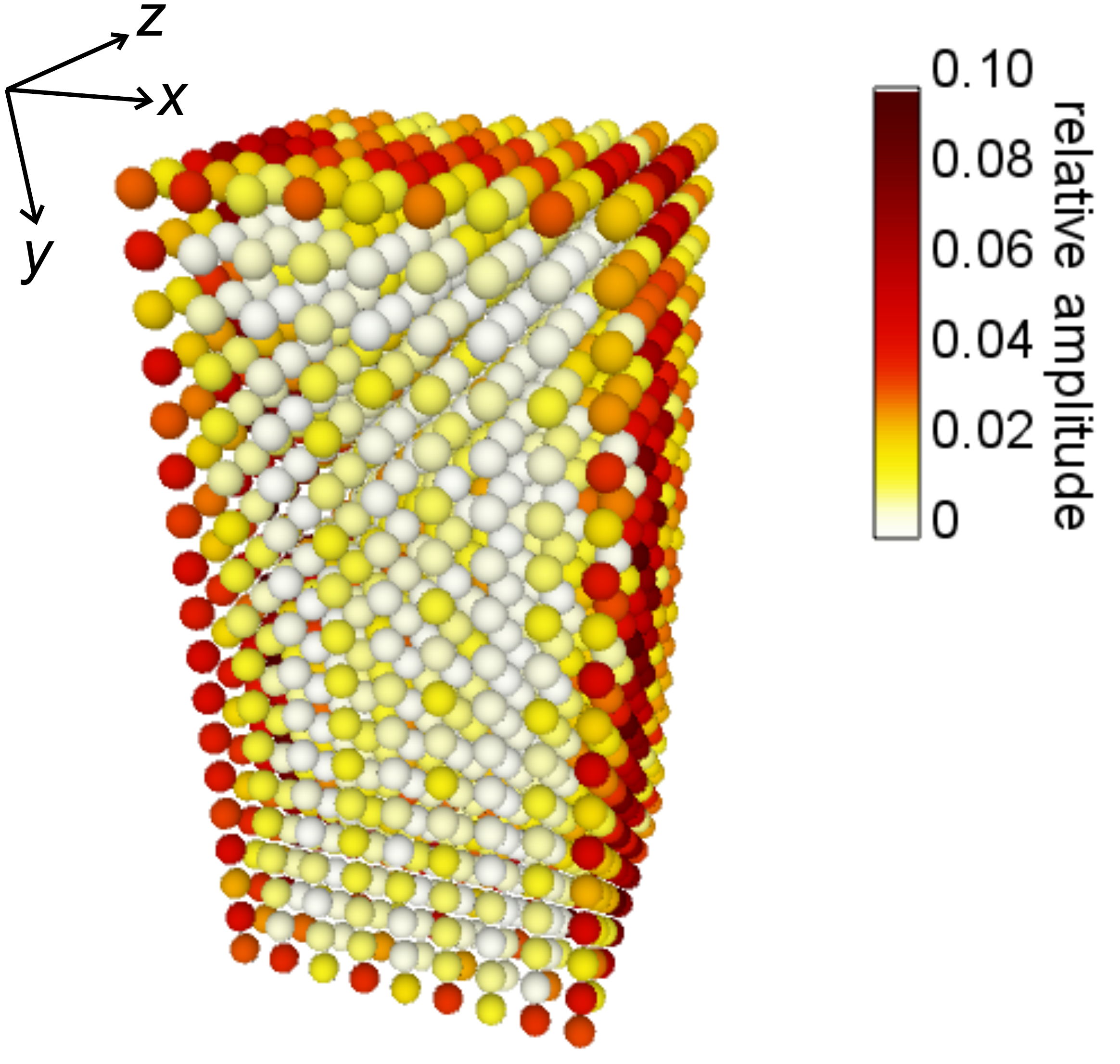}
\caption{\label{fig7} Single-shot image of time evolution of the proposed photonic system. The color of each sphere represents the relative amplitude at each site, which is normalized so that the sum of squares is unity. One can confirm the lasing of boundary modes. The parameters used are $a_x=a_y=1$, $a_z=0.1$, and $\beta=0.5$.
}
\end{figure}

To be concrete, we consider aligning ring resonators on lattice points of a square lattice to construct an analog of a quantum spin Hall system. We can realize artificial gauge fields (i.e., phases gained through hoppings) by tuning the length of waveguides between resonators. The dynamics of such a resonator system is effectively described by the Hamiltonian of the Harper-Hofstadter model, which imitates electrons in a square lattice under the uniform magnetic field \cite{Hofstadter1976,Hafezi2011,Hafezi2013,Ozawa2019}. We note that the layers depicted in Fig.~\ref{fig5}a do {\it not} correspond to the layers of quantum spin Hall systems, while ring resonators on a plane parallel to the $xy$ plane form the counterpart of a quantum spin Hall system. To realize non-Hermitian coupling between the effective spins, we propose to use waveguide segments with mirrors and homogeneous gain which are studied in previous research \cite{Fan2003,Song2020}. It is noteworthy that we need to carefully engineer the effective gauge fields to only amplify the surface modes. For this purpose, tuning the length of the waveguide segments, one should change the phase gained through the wave propagation in the waveguides in each layer. In addition, signs of the phases gained through hoppings are alternate between odd- and even-numbered rows, which realizes the interlayer interaction between different effective spin degrees of freedom.

Altogether, we obtain the optical system whose dynamics is described by the following effective Hamiltonian,
\begin{eqnarray}
 H &=& \sum_{\mathbf{r},s = \pm}  (a_x e^{is\phi y} \hat{c}^{\dagger}_{\mathbf{r},s} \hat{c}_{\mathbf{r}+\mathbf{e}_x,s} + a_y \hat{c}^{\dagger}_{\mathbf{r},s} \hat{c}_{\mathbf{r}+\mathbf{e}_y,s}) + {\rm H.c.} \nonumber\\
  &{}& + \sum_{\mathbf{r}} [a_z \sin(\phi y) (\hat{c}^{\dagger}_{\mathbf{r},+} \hat{c}_{\mathbf{r}+\mathbf{e}_z,-} + \hat{c}^{\dagger}_{\mathbf{r},-} \hat{c}_{\mathbf{r}+\mathbf{e}_z,+}) + {\rm H.c.} \nonumber\\
  &{}& + i\beta \sin(\phi y) (\hat{c}^{\dagger}_{\mathbf{r},+} \hat{c}_{\mathbf{r},-} + \hat{c}^{\dagger}_{\mathbf{r},-} \hat{c}_{\mathbf{r},+})] , \label{Hamiltonian_optical}
\end{eqnarray}
which is derived from the coupled-mode theory \cite{Haus1991}. $\hat{c}^{\dagger}_{\mathbf{r},s}$, $\hat{c}_{\mathbf{r},s}$ represent the bosonic creation and annihilation operators at the location $\mathbf{r}=(x,y,z)$ and the effective spin $s=\pm$. In addition, $\mathbf{e}_i$ is the unit vector directed to the $i$ direction. $\phi$ is the phase parameter and set to be $\phi=2\pi/3$ in the proposed optical system since we periodically align three different layers of ring-resonator arrays. This effective Hamiltonian exhibits lasing surface modes protected by EPs, which share the same properties as demonstrated for a prototypical model above.

We numerically demonstrate the existence of lasing surface modes in the effective Hamiltonian of the proposed ring-resonator array \eqref{Hamiltonian_optical}. First, we perform the band calculation of the effective Hamiltonian. We impose the open boundary condition in the $x$ direction and the periodic boundary conditions in the $y$ and $z$ directions. Figure \ref{fig6} shows the band structure of the surface modes. As in our prototypical model (Eq.~\eqref{prototypical_model} and Fig.~\ref{fig1}), we can confirm the existence of gapless modes with larger imaginary parts of eigenvalues than those of the bulk modes. The gapless surface modes also exhibit four EPs in their surface bands. Therefore, one can expect that the proposed ring-resonator array exhibits lasing surface modes that are protected by EPs and robust against disorder.

We also demonstrate the dynamical emergence of lasing surface modes in the proposed optical system by numerically simulating its time evolution. We consider the fully open boundary conditions (i.e., the open boundary conditions in the $x$, $y$, $z$ directions) and perform the fourth-order Runge-Kutta simulation as in Fig.~\ref{fig3}(b). We set the time step as $dt = 0.001$ and use a $9\times18\times9$ cubic lattice. We consider the random initial condition, where the amplitude of each site is determined so that the real and imaginary parts follow a uniform distribution ranging $[-0.01,0.01]$ for each. Figure \ref{fig7} shows the snapshot of the relative-amplitude distribution obtained from the numerical calculation. We confirm the emergence of boundary-localized modes at the boundaries perpendicular to the stacked quantum spin Hall layers. The amplitudes of the surface sites are amplified while those of the bulk sites are suppressed, which indicates the lasing of exceptional surface modes.

\begin{figure}[t]
  \includegraphics[width=60mm,bb=0 0 340 340,clip]{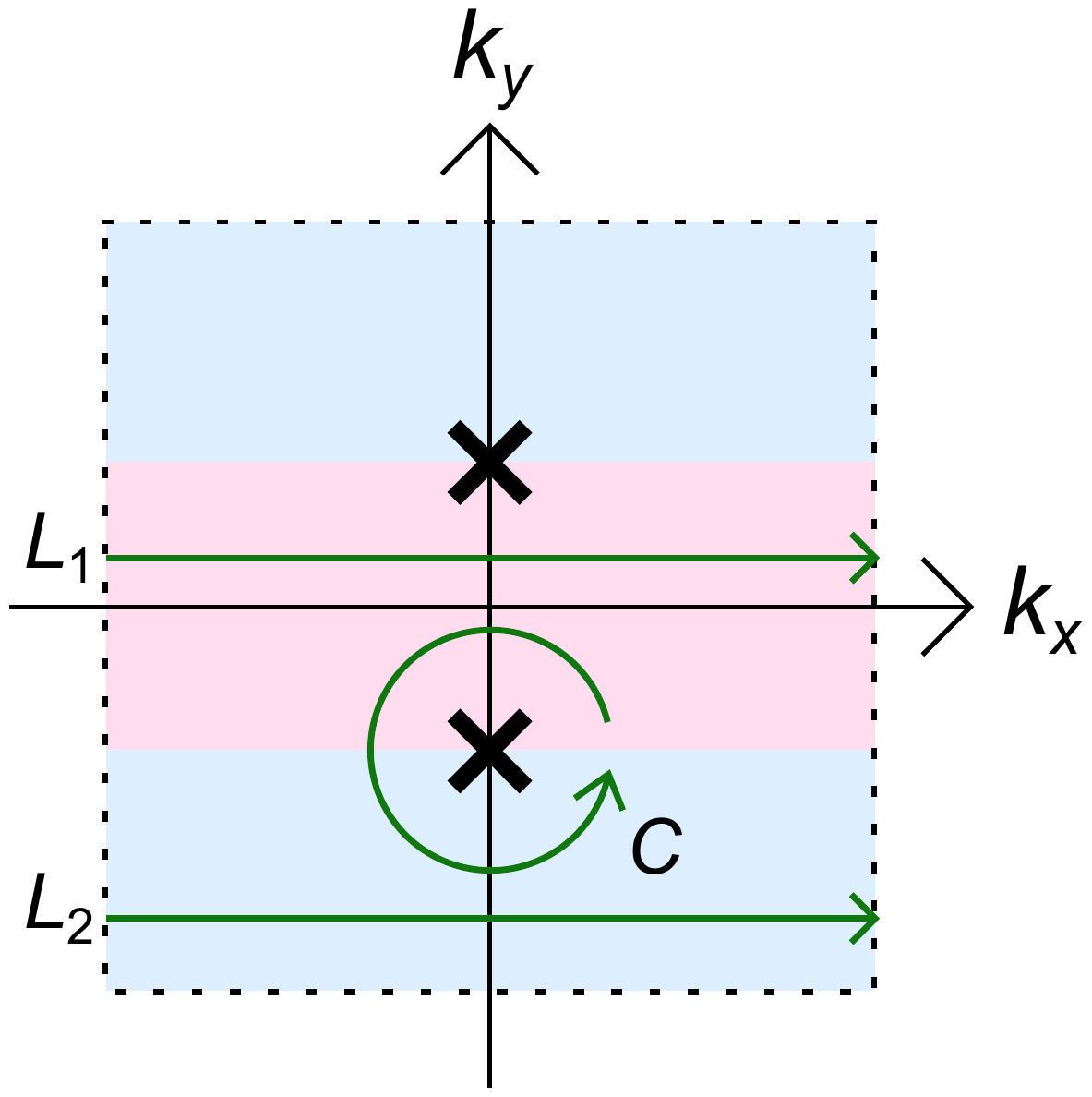}
  \caption{\label{fig8} Integral paths of winding numbers in the band structure with one pair of EPs. The crosses represent the EPs in the two-dimensional (generalized) Brillouin zone. The green arrows show the integral paths considered in the argument of the no-go theorem. The blue and red regions represent the difference in the winding numbers obtained from the integral paths in those regions.
}
\end{figure}

\section{No-go theorem of surface modes with a single pair of EPs\label{section4}}
We reveal that four EPs in the dispersion relations of lasing surface modes are essential to realize robust gapless modes protected by those EPs. In fact, we can theoretically show that if a two-dimensional band, such as the surface band of a topological laser, has only a pair of EPs under the periodic boundary conditions, they coalesce and disappear under the fully open boundary conditions. In addition, the non-Hermitian doubling theorem \cite{Yang2021,Bessho2021} prohibits the appearance of an odd number of EPs \footnote{Strictly, the doubling theorem does not prohibit the appearance of an odd number of exceptional points on a single surface of a three-dimensional system. However, if we consider both sides of the system, exceptional points should appear in pairs and our argument should be valid in such cases.}. Therefore, four or more EPs are necessary to make sure that EPs robustly remain under the fully open boundary conditions.

This point can be inferred from the fact that a single pair of EPs accompany nonzero winding numbers which induce the non-Hermitian skin effect \cite{Martinez2018,Xiong2018,Yao2018,Yokomizo2019,Okuma2020,Zhang2020}. Since such nonzero winding number is prohibited under the open boundary conditions, a pair of EPs also disappear. Specifically, we consider two integral paths $L_{1,2}$ of winding numbers in Fig.~\ref{fig8}. The winding number $w = (1/2\pi i) \int \partial_\mathbf{k} \log \det H(\mathbf{k}) d\mathbf{k}$ is a topological invariant and thus unchanged under the deformation of the integral path until the path crosses EPs. Therefore, the difference of the winding numbers on $L_{1,2}$ is equal to that on the circular integral path $C$ around the EP,
\begin{eqnarray}
w_{L_1} - w_{L_2} = \frac{1}{2\pi i} \oint_{C} \partial_\mathbf{k} \log \det H(\mathbf{k}) d\mathbf{k} = w_{C}. \label{winding_numbers}
\end{eqnarray}
If $|w_C|$ becomes two or more, the EP can split unless symmetries protect their degeneracy \cite{Yang2021,Delplace2021}. To omit the cases that such degenerate EPs appear and thus the numbers of EPs are effectively more than two, we assume $|w_C| = 1$. Then, Eq.~\eqref{winding_numbers} indicates that at least either one of the winding numbers $w_{L_1}$, $w_{L_2}$ is nonzero and even if both are nonzero, they have the same sign. If we regard the system with open boundaries as a one-dimensional system with long unit cells extended to the $y$ direction, the winding number is equal to the integral of the winding numbers obtained from integral paths parallel to $L_1$ and $L_2$, which are equal to $w_{L_1}$ or $w_{L_2}$. Since $w_{L_1}$ and/or $w_{L_2}$ are nonzero, such winding number of a large one-dimensional system is also nonzero, which leads to the non-Hermitian skin effect. The non-Hermitian skin effect drastically modifies the spectrum so that the winding number of the open boundary spectrum becomes zero \cite{Okuma2020,Zhang2020}. Therefore, a single pair of EPs are prohibited under the fully open boundary conditions.

These arguments imply that our model \eqref{prototypical_model} is a minimal one because robust exceptional surface modes require at least four EPs. While we can construct a model of exceptional surface modes that exhibit two EPs under the periodic boundary conditions, such surface modes open a gap under the open boundary conditions due to the non-Hermitian skin effect (cf.~Appendix \ref{appx_D}). We note that while the theory of the non-Hermitian skin effect in two-dimensional systems is not fully established, our arguments are based on the analysis of (effectively) one-dimensional systems.

\section{Summary and Discussions\label{section5}}
We proposed a three-dimensional topological surface laser utilizing exceptional surface modes protected by exceptional points (EPs). We constructed such topological insulator laser by introducing a non-Hermitian coupling to a weak topological insulator. We also proposed a ring-resonator realization of the topological surface laser. These results indicate that exceptional surface modes serve as a general guiding principle to construct topological lasers and enable their flexible design. In addition, we showed that the number of EPs must be four or more to make sure that EPs stably exist under the open boundary conditions, as is the case in the proposed topological surface laser.

The proposed topological laser is stable without symmetry protections and judicious gains at the edge of the samples, which have been prerequisites for two-dimensional topological lasers \cite{St-Jean2017,Harari2018,Bandres2018,Song2020,Sone2020}. Therefore, it is robust against symmetry-breaking noises and distortion of surface configurations, which will expand the potential of non-Hermitian devices. Some two-dimensional topological lasers have required symmetry protections because EPs in their edge modes require symmetries, such as the $PT$ symmetry \cite{Kawabata2019b,Sone2020}. Meanwhile, the three-dimensional topological laser requires no symmetries since EPs in the surface bands can be topologically protected without symmetries. In addition, since the complex dispersions of the surface modes exhibit positive imaginary parts of the eigenvalues, they are amplified without additional gains and thus without the need for the knowledge of the surface configuration. While a previous study \cite{Denner2021} has also analyzed symmetry-free topological modes in three-dimensional systems, such topological modes accompany nontrivial winding numbers of complex bands and their eigenvalues appear inside of the bulk bands. On the contrary, our proposal enables imaginary parts of the eigenvalues of the surface modes to be larger than those of bulk ones, which is essential to realize topological lasers. We also note that topological systems analyzed in Ref.~\cite{Denner2021} are termed as exceptional topological insulators, but these are radically different from the exceptional boundary modes analyzed in the present study. This is because unless the exceptional topological insulators, the exceptional boundary modes utilize the topology of boundary bands and are independent of the bulk topology.

On the one hand, gapless boundary modes protected by EPs indicate the breakdown of the bulk-boundary correspondence in non-Hermitian systems \cite{Sone2020}. Our results in three-dimensional systems imply that the breakdown of the bulk-boundary correspondence in this sense is common in non-Hermitian systems. On the other hand, our argument revealed that EPs can disappear under the open boundary conditions due to the non-Hermitian skin effect. Therefore, it is desirable to construct a classification of non-Hermitian topology reflecting both the exceptional boundary modes and the skin effect.

While we proposed an optical setup of the topological surface laser, it has remained an intriguing problem whether or not we can realize exceptional surface modes and lasing devices in other physical systems. Especially, previous research has shown that exceptional edge modes in two-dimensional systems can be realized in active matter \cite{Sone2020}. Therefore, the extension to three-dimensional topological active matter is worth exploring. It is also of practical interest to study if the exceptional surface modes can exhibit EP-inducing unique phenomena, such as interchanging of eigenstates \cite{Dembowski2001} and enhanced sensitivity \cite{Liu2016,Chen2017}.

\begin{acknowledgments}
We thank Eiji Saitoh and Taro Sawada for valuable discussions. K.S. is supported by World-leading Innovative Graduate Study Program for Materials Research, Industry, and Technology (MERIT-WINGS) of the University of Tokyo. K.S. is also supported by JSPS KAKENHI Grant Number JP21J20199. Y.A. is supported by JSPS KAKENHI Grant Number JP19K23424. T.S. is supported by JSPS KAKENHI Grant Numbers JP19H05796 and JST, CREST Grant Number JPMJCR20C1, Japan. T.S. is also supported by Institute of AI and Beyond of the University of Tokyo.
\end{acknowledgments}

\appendix

\begin{figure}[b]
  \includegraphics[width=70mm,bb=0 0 510 360,clip]{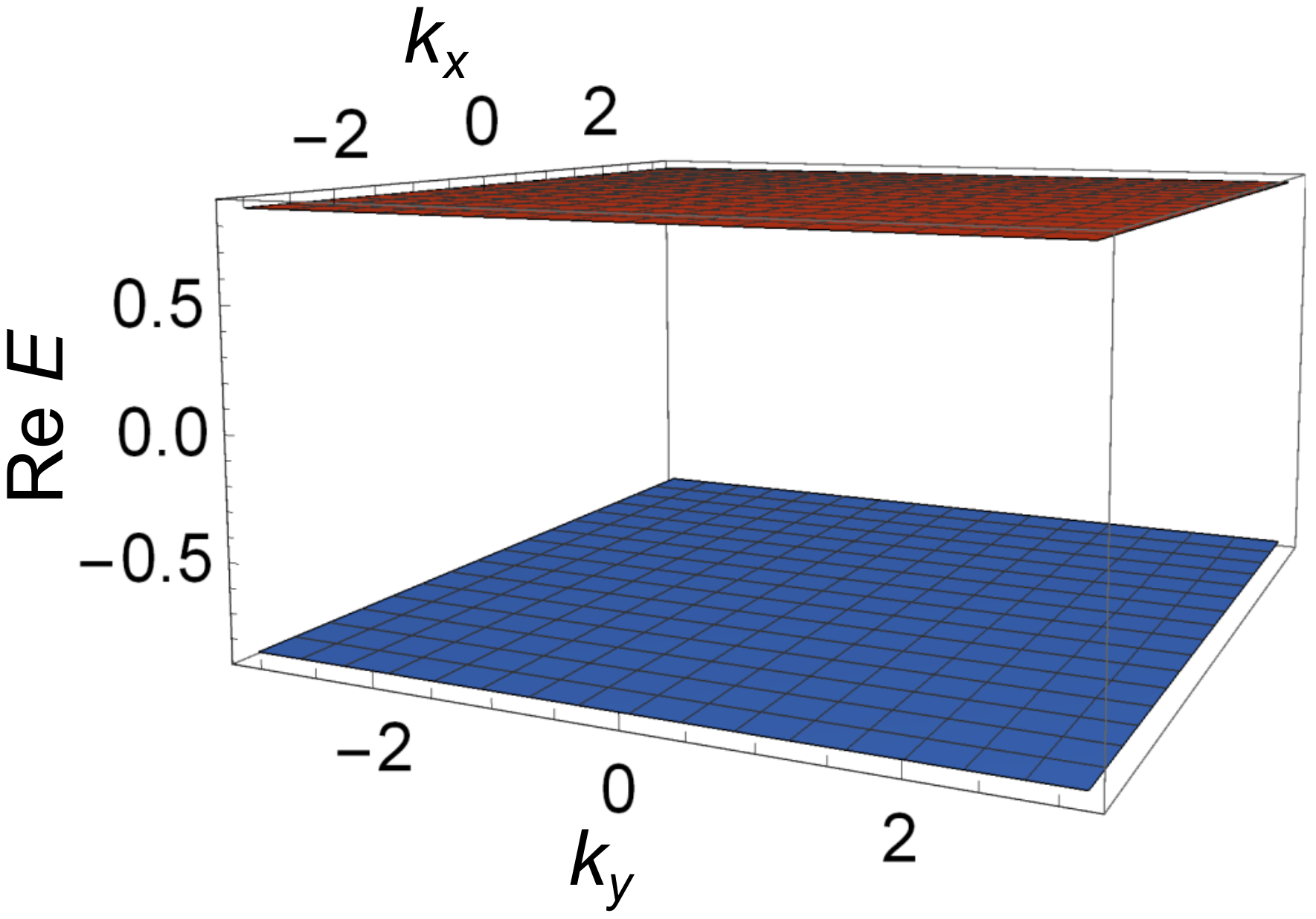}
\caption{\label{fig9} Band structure of the Hamiltonian of our model under the open boundary condition in the $z$ direction. We only show the bands up to the fourth closest to the ${\rm Re}\,E=0$ plane. The parameters used are $u=-1$, $c=0.3$, $\gamma=0.5$, and there are no disorder terms.}
\end{figure}

\section{\label{appx_A}Real-space descriptions of the Hamiltonians of the proposed models.}
Here, we explicitly provide the real-space description of our models in the main text. Since the wavenumber-space description is obtained via the Fourier transformation of the Hamiltonian in the real space, we can recover the real-space description by substituting $\cos k_j$ and $\sin k_j$ into $\sum_{\mathbf{r}}(|\mathbf{r}\rangle\langle \mathbf{r}+\mathbf{e}_j|+|\mathbf{r}+\mathbf{e}_j\rangle\langle \mathbf{r}|)/2$ and $\sum_{\mathbf{r}}(-i|\mathbf{r}\rangle\langle \mathbf{r}+\mathbf{e}_j|+i|\mathbf{r}+\mathbf{e}_j\rangle\langle \mathbf{r}|)/2$ ($\mathbf{r}$ is the location and $\mathbf{e}_j$ is the unit vector directed to the $j$ direction) for each.

The Hamiltonian of the prototypical model (Eq.~\eqref{prototypical_model} in the main text) is described in the real space as
\begin{eqnarray}
 H &=& \sum_{\mathbf{r}}\left\{ \left[u|\mathbf{r}\rangle\langle \mathbf{r}|+ \frac{1}{2}(|\mathbf{r}\rangle\langle \mathbf{r}+\mathbf{e}_x|+|\mathbf{r}+\mathbf{e}_x\rangle\langle \mathbf{r}| \right. \right.  \nonumber\\
 &{}& \left. \left. + |\mathbf{r}\rangle\langle \mathbf{r}+\mathbf{e}_y|+|\mathbf{r}+\mathbf{e}_y\rangle\langle \mathbf{r}|) \right] I \otimes \sigma_z \right. \nonumber\\
 &{}&-\frac{i}{2} (|\mathbf{r}\rangle\langle \mathbf{r}+\mathbf{e}_x|-|\mathbf{r}+\mathbf{e}_x\rangle\langle \mathbf{r}|)\sigma_z \otimes \sigma_x  \nonumber\\
 &{}& -\frac{i}{2} (|\mathbf{r}\rangle\langle \mathbf{r}+\mathbf{e}_y|-|\mathbf{r}+\mathbf{e}_y\rangle\langle \mathbf{r}|) I \otimes \sigma_y \nonumber\\
 &{}& \left. + \left[\frac{c}{2}(|\mathbf{r}\rangle\langle \mathbf{r}+\mathbf{e}_z|+|\mathbf{r}+\mathbf{e}_z\rangle\langle \mathbf{r}|)+i\gamma|\mathbf{r}\rangle\langle \mathbf{r}|\right] \sigma_y \otimes \sigma_x \right\},\nonumber\\ \label{prototypical_model_realspace}
\end{eqnarray}
where $\sigma_i$ represents the Pauri matrices and $I$ is the $2\times2$ identity matrix. The parameters are the same as in Eq.~\eqref{prototypical_model} in the main text. To realize the open boundary condition, we assume that the amplitude is zero at $\mathbf{r}$ corresponding to the outside of the system. We also construct the modified model (Eq.~\eqref{propagating_surface_modes} in the main text) to show the existence of propagating lasing surface modes. Its Hamiltonian is described in the real space as
\begin{eqnarray}
 H(\mathbf{k}) &=& \sum_{\mathbf{r}}\left\{ \left[u|\mathbf{r}\rangle\langle \mathbf{r}|+ \frac{1}{2}(|\mathbf{r}\rangle\langle \mathbf{r}+\mathbf{e}_x|+|\mathbf{r}+\mathbf{e}_x\rangle\langle \mathbf{r}| \right. \right.  \nonumber\\
 &{}& \left. \left. + |\mathbf{r}\rangle\langle \mathbf{r}+\mathbf{e}_y|+|\mathbf{r}+\mathbf{e}_y\rangle\langle \mathbf{r}|) \right] \left( a I \otimes \sigma_z + b \sigma_z \otimes \sigma_z \right) \right.\nonumber\\
 &{}& -\frac{i}{2} (|\mathbf{r}\rangle\langle \mathbf{r}+\mathbf{e}_x|-|\mathbf{r}+\mathbf{e}_x\rangle\langle \mathbf{r}|) \left( b I \otimes \sigma_x + a \sigma_z \otimes \sigma_x \right) \nonumber\\
 &{}& -\frac{i}{2} (|\mathbf{r}\rangle\langle \mathbf{r}+\mathbf{e}_y|-|\mathbf{r}+\mathbf{e}_y\rangle\langle \mathbf{r}|) \left( a I \otimes \sigma_y + b \sigma_z \otimes \sigma_y \right) \nonumber\\
 &{}& \left. + \left[\frac{c}{2}(|\mathbf{r}\rangle\langle \mathbf{r}+\mathbf{e}_z|+|\mathbf{r}+\mathbf{e}_z\rangle\langle \mathbf{r}|)+i\gamma|\mathbf{r}\rangle\langle \mathbf{r}|\right] \sigma_y \otimes \sigma_x \right\}. \nonumber\\ \label{propagating_surface_modes_realspace}
\end{eqnarray}

\begin{figure*}[t]
  \includegraphics[width=160mm,bb=0 0 1590 710,clip]{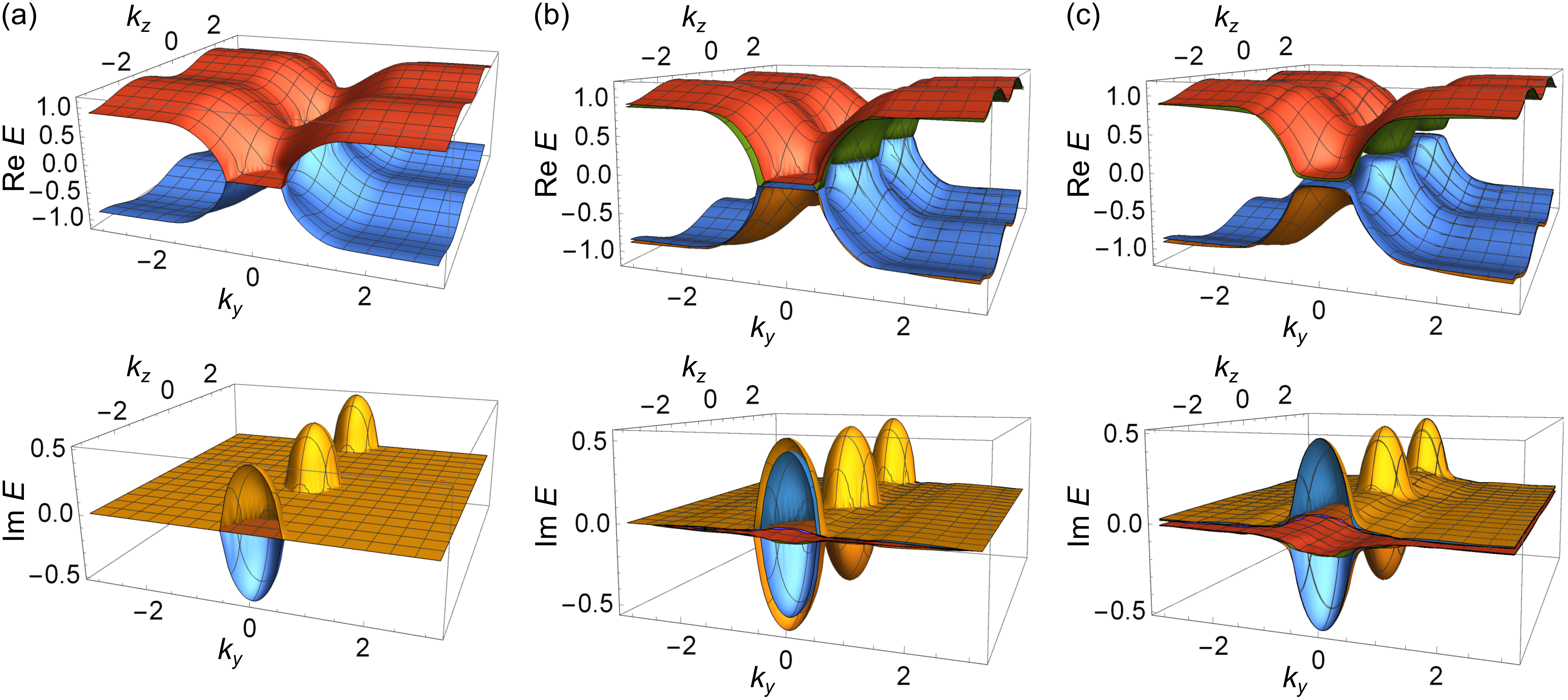}
\caption{\label{fig10} Band structure of exceptional surface modes with exceptional rings protected by symmetries. For clarity, we only present the bands up to the fourth closest to the ${\rm Re}\,E=0$ plane in the upper panels. In the lower panels, we show eight bands with up to the fourth largest and smallest imaginary parts of eigenvalues. (a) Exceptional surface modes in the system without disorder. There exist gapless modes with two exceptional rings. The parameter used are $u=-1$, $c=0.7$, and $\beta=0.5$. (b) Robustness of the exceptional surface modes against an on-site random potential and an imaginary random spin coupling. The noise intensity is set to be $W_u=0.1$, $W_a=0$, and $W_b=0.1$ and the other parameters are the same as in panel (a). The exceptional rings and the gapless modes robustly remain even under the existence of such disorders. (c) Instability of the exceptional surface modes against a real random spin coupling. The noise intensity is set to be $W_u=0.1$, $W_a=0.2$, and $W_b=0$ and the other parameters are the same as in panel (a). In this case, exceptional rings disappear and thus the bands open a gap.
}
\end{figure*}

\section{\label{appx_B}Absence of gapless modes in the surface of the proposed topological surface laser perpendicular to the $z$ axis}

As discussed in the main text, the proposed topological laser exhibits lasing surface modes only at the boundaries perpendicular to the stacked layers. Here, we confirm such anisotropy by the numerical calculation of the band structure under the open boundary condition in the $z$ direction. We consider the prototypical model (Eq.~\eqref{prototypical_model} in the main text) of the topological surface laser. We impose the periodic boundary conditions in the $x$ and $y$ directions. We use the same parameters ($u=-1$, $c=0.3$, $\gamma=0.5$) as in Fig.~\ref{fig1} and consider no disorder terms.

Figure \ref{fig9} presents the numerical result of the band calculation. One can confirm the absence of gapless modes under the open boundary condition only in the $z$ direction. Therefore, the lasing surface modes disappear at the boundary in the $z$ direction, which is also confirmed in Fig.~\ref{fig3}(b) in the main text. We note that similar anisotropy can emerge in a weak topological insulator \cite{Fu2007,Ringel2012}, which we utilize to construct the proposed topological laser.

\section{\label{appx_C}Exceptional surface modes exhibiting exceptional rings}

While we propose and analyze a model of lasing surface modes protected by four exceptional points (EPs), we can construct exceptional surface modes that exhibit exceptional lines or exceptional rings in their surface bands. Such exceptional surface modes can be realized in the following non-Hermitian Hamiltonian,
\begin{eqnarray}
 H(\mathbf{k}) &=& (u+\cos k_x + \cos k_y) I \otimes \sigma_z + \sin k_x \sigma_z \otimes \sigma_x \nonumber\\
 &{}& + \sin k_y I \otimes \sigma_y + c\sin k_z \sigma_y \otimes \sigma_x  + i\beta \sigma_x \otimes \sigma_x, \nonumber\\ \label{exceptional_ring}
\end{eqnarray}
with $u$, $c$, $\beta$ being real parameters. As in Eq.~\eqref{prototypical_model} in the main text, $\sigma_i$ and $I$ represent the Pauri matrices and the $2\times2$ identity matrix respectively. We note that the non-Hermitian term, $i\beta \sigma_x \otimes \sigma_x$, is different from that in our prototypical model analyzed in the main text. 

We calculate the surface band structure of this Hamiltonian and confirm the emergence of exceptional surface modes with exceptional nodal lines. We impose the open boundary condition in the $x$ direction and the periodic boundary conditions in the $y$ and $z$ directions. We use the parameters $u=-1$, $c=0.7$, and $\beta=0.5$. As shown in Fig.~\ref{fig10}(a), exceptional rings appear in the band of the surface modes when we set $c$ to be larger than $\beta$. These exceptional rings protect the gapless modes from opening the gap against disorder.

However, the topological protection of exceptional rings in two-dimensional bands (e.g. the surface bands in Fig.~\ref{fig10}) requires $\mathbb{Z}_2$ symmetries, such as $PT$ symmetry, $CP$ symmetry, pseudo-Hermiticity, and chiral symmetry \cite{Kawabata2019b}. Therefore, disorder breaking these symmetries can lift a gap in the surface bands \cite{Sone2020}. We confirm the robustness and instability of the exceptional surface modes against symmetry-preserving and -breaking noises. Specifically, we consider the random real on-site potential, i.e., real diagonal terms whose intensities follow uniform distributions ranging $[-W_{u},W_{u}]$. We also utilize the complex random spin coupling, which couples the first and fourth components or the second and third ones and has the random complex intensity whose real and imaginary parts follow uniform distributions ranging $[-W_{a,b},W_{a,b}]$ respectively. In Fig.~\ref{fig10}(b), we introduce the random on-site potential and the imaginary spin coupling. Since such disorders preserve the $PT$ symmetry, the exceptional surface modes robustly exist. On the other hand, we introduce the real random spin coupling in Fig.~\ref{fig10}(c), which breaks the $\mathbb{Z}_2$ symmetries in the present model. We confirm that gapless modes disappear in this case, which implies the instability of the exceptional surface modes against symmetry-breaking disorder. We note that the lasing surface modes with EPs are robust against the disorders considered in this section because EPs are topologically protected even without symmetries.

\section{\label{appx_D}Exceptional surface modes with a single pair of EPs and its absence under the fully open boundary conditions}

In the main text, we show that a single pair of EPs are unstable and disappear under the fully open boundary conditions. While we can construct a model of exceptional surface modes that exhibits a single pair of EPs under the open boundary condition only in one direction, such gapless modes disappear due to the non-Hermitian skin effect \cite{Martinez2018,Xiong2018,Yao2018,Yokomizo2019,Okuma2020,Zhang2020} under the fully open boundary conditions. Specifically, we consider a toy model described by the following Hamiltonian,
\begin{eqnarray}
 H(\mathbf{k}) &=& (u+\cos k_x + \cos k_y + \cos k_z) I \otimes \sigma_z \nonumber\\
 &{}& + (\sin k_x + ib_x) \sigma_z \otimes \sigma_x  + (\sin k_y+ib_y) I \otimes \sigma_y \nonumber\\
 &{}& + (\sin k_z+ib_z) \sigma_y \otimes \sigma_x + \alpha \sigma_x \otimes \sigma_x,\label{two_EP}
\end{eqnarray}
where $\sigma_i$ and $I$ represent the Pauri matrices and the $2\times2$ identity matrix respectively. This Hamiltonian is constructed from the model of a topological insulator with the non-Hermitian terms proportional to $b_i$ ($i=x,y,z$). The term proportional to $\alpha$ breaks the time-reversal symmetry, which is essential for the protection of gapless modes in a topological insulator. In the following numerical calculations, we focus on the parameters $u=-2$, $b_x=b_y=0$, $b_z=0.5$, and $\alpha=0.1$.

\begin{figure}[t]
  \includegraphics[width=60mm,bb=0 0 510 730,clip]{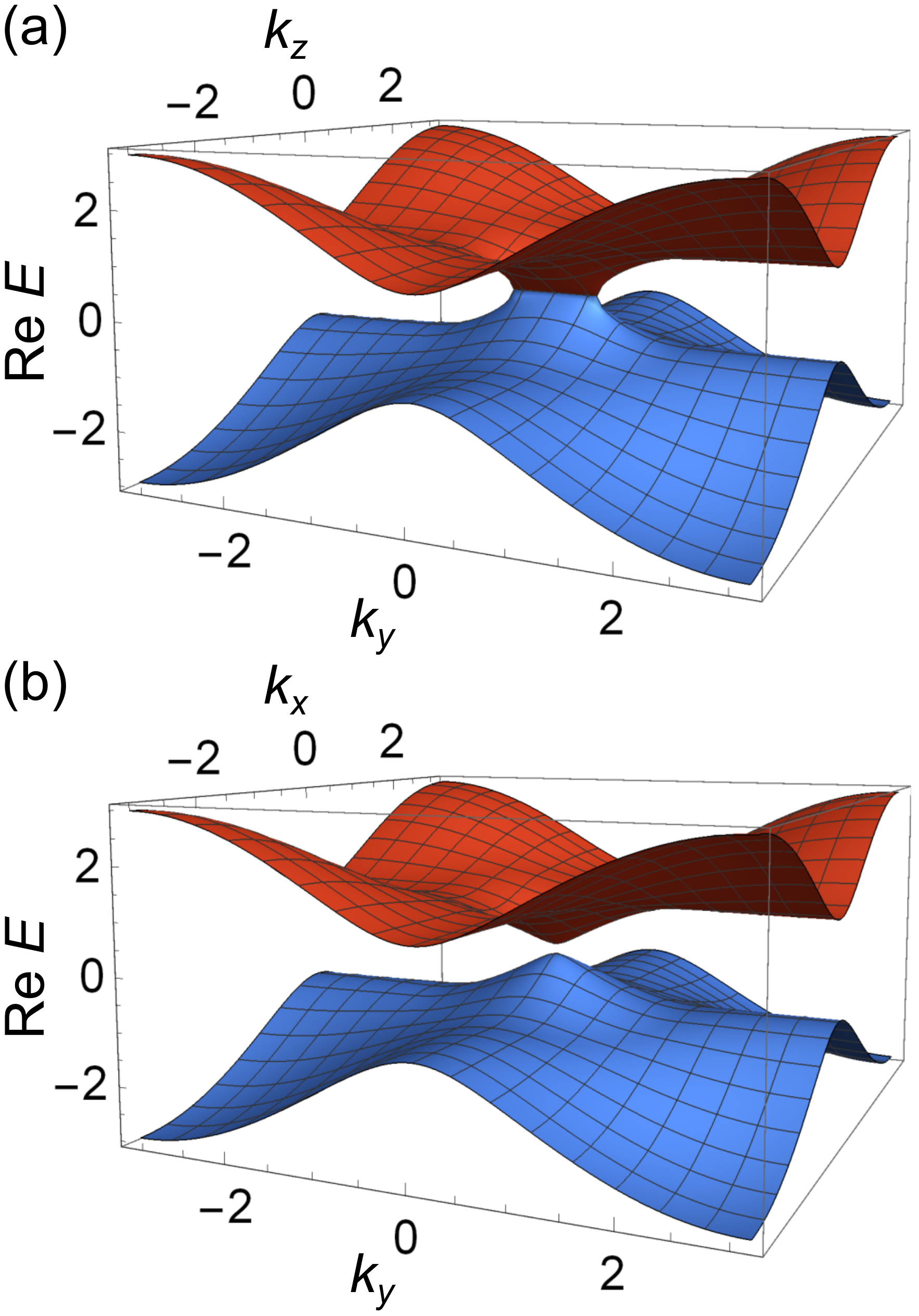}
\caption{\label{fig11} Exceptional surface modes with a pair of EPs. (a) Surface band structure under the open boundary condition in the $x$ direction. We show the real part of eigenvalues. There are gapless modes with a single pair of EPs. The parameters used are $u=-2$, $b_x=b_y=0$, $b_z=0.5$, and $\alpha=0.1$. (b) Surface band structure under the open boundary condition in the $z$ direction. In contrast to panel (a), gapless modes are absent under this boundary condition. The parameters used are the same as in panel (a).
}
\end{figure}

First, we calculate the surface band structure under the open boundary condition in the $x$ direction and the periodic boundary conditions in the $y$ and $z$ directions to confirm the existence of gapless modes with EPs under such boundary conditions. Figure \ref{fig11}(a) shows the numerical result. We obtain gapless modes that exhibit a single pair of EPs in their surface bands. While under the periodic boundary condition, the EPs can protect the gapless modes against disorder, they are unstable due to the skin effect under the fully open boundary conditions. In addition, the skin effect also affects the surface band under the open boundary condition in the $z$ direction and the periodic boundary conditions in the $x$ and $y$ directions. Under such boundary conditions, gapless modes disappear as shown in Fig.~\ref{fig11}(b).

\begin{figure}[t]
  \includegraphics[width=60mm,bb=0 0 470 680,clip]{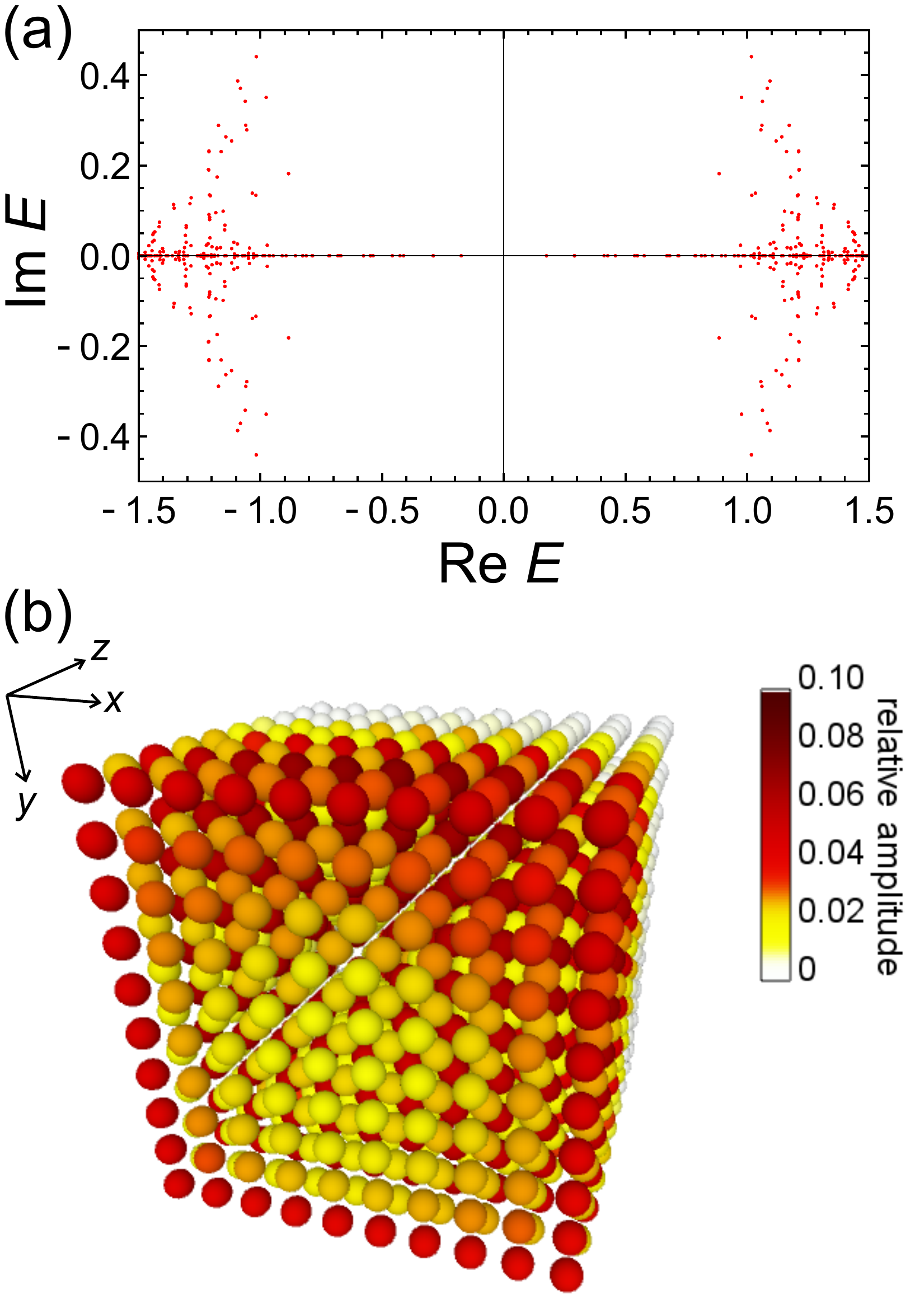}
\caption{\label{fig12} Absence of exceptional surface modes under the fully open boundary conditions. (a) Eigenvalue distribution under the fully open boundary conditions. Red dots show the eigenvalues of the Hamiltonian \eqref{two_EP} in a $10\times10\times10$ cubic lattice. There are no gapless modes around the ${\rm Re}\,E=0$ axis. We use the same parameter as in Fig.~\ref{fig11}. (b) Amplitude distribution obtained from the numerical simulation of the dynamics. The color of each sphere represents the relative amplitude at each site, which is normalized so that the sum of squares is unity. The sites with small $z$ coordinates exhibit large amplitiudes, which indicate the localization of the bulk modes. The parameters used are the same as in panel (a).
}
\end{figure}

We next confirm the disappearance of the gapless modes in this model \eqref{two_EP} under the fully open boundary conditions. We calculate the eigenvalues of the Hamiltonian in a $10\times10\times10$ cubic lattice with open boundaries. Figure \ref{fig12}(a) presents the spectrum of the Hamiltonian. One can confirm the absence of the surface modes around the ${\rm Re}\,E=0$ axis. Therefore, we can conclude that the exceptional surface modes seen in Fig.~\ref{fig11} are unstable under the fully open boundary conditions. We also confirm the emergence of the non-Hermitian skin effect in this model by calculating the dynamics of this Hamiltonian. Specifically, we perform the Runge-Kutta simulation as in Figs.~\ref{fig3}, \ref{fig7} with the time step being $dt = 0.001$. We consider the random initial condition, where the amplitude of each site is determined so that the real and imaginary parts follow a uniform distribution ranging $[-1,1]$ for each. We obtain the localized steady state shown in Fig.~\ref{fig12}(b), which indicates the emergence of the non-Hermitian skin effect.

The non-Hermitian skin effect modifies the spectrum of the Hamiltonian \eqref{two_EP} and eliminates the exceptional surface modes under the fully open boundary conditions. Here, we analytically show the disappearance of EPs and its relation to the non-Hermitian skin effect. To predict the behavior under the open boundary condition, we use the non-Bloch band theory developed in a recent study \cite{Yokomizo2019}. In the non-Bloch band theory, we consider the extension of wavenumbers to complex ones, which properly describes the effect of localization by the non-Hermitian skin effect. Then, we obtain the non-Bloch Hamiltonian of our model \eqref{two_EP} and reveal that it can be described as the transformation of a Hermitian Hamiltonian by a regular matrix. Thus, exceptional surface modes must be absent in the model under the fully open boundary conditions.

First, we describe the procedure to determine the imaginary parts of wavenumbers (i.e., the generalized Brillouin zone) in the non-Bloch band theory \cite{Yokomizo2019}. In a discrete periodic system, the Bloch Hamiltonian $H(k)$ can be regarded as the matrix function of $\beta=e^{ik}$. Then, we extend the domain of definition of $H(\beta)$ to the entire complex plane. The eigenequation of this extended Bloch Hamiltonian reads $\det(H(\beta)-EI)=0$ and has $2M$ answers $\beta_1,\cdots,\beta_{2M}$ for each fixed $E$ with $M$ being the product of the hopping range and the internal degrees of freedom. The generalized Brillouin zone and its spectrum are determined as the pairs of $\beta_M$ and $E$ that satisfy $|\beta_M|=|\beta_{M+1}|$. 

Since the non-Bloch band theory is established only in one-dimensional systems, we regard the Hamiltonian of our model \eqref{two_EP} as that of a one-dimensional periodic system by fixing the wavenumbers in the $x$ and $y$ direction, $k_x$, $k_y$. We also assume $b_x=b_y=0$ as in Figs.~\ref{fig11}, \ref{fig12}. Then, we obtain the effectively one-dimensional Bloch Hamiltonian 
\begin{eqnarray}
 H(k_x) = (u'+\cos k_z) \gamma_1 + (\sin k_z+ib_z) \gamma_2+c\gamma_3,\ \  \label{one-dimension}
\end{eqnarray}
with $u'$ and $c$ being $u'=u+\cos k_x+\cos k_y$, $c=\sqrt{\sin^2 k_x+\sin^2 k_y+\alpha^2}$. We note that $\gamma_i$ represents the gamma matrices and we can substitute them into the Pauri matrices $\gamma_i=\sigma_i$ for the purpose of investigating the non-Bloch band.

To analytically perform the calculation of the generalized Brillouin zone, we rewrite $\beta_M=\beta_{M+1}e^{-i\theta}=\beta$ and consider the equation $\det(H(\beta)-EI) - \det(H(\beta e^{i\theta})-EI) = 0$ as discussed in \cite{Yokomizo2019}. The equation reads
\begin{equation}
 (u'+b_z)(e^{i\theta}-1)\beta-(u'-b_z)(1-e^{-i\theta})\beta^{-1} = 0,\label{beta_eq1}
\end{equation}
which is independent of $E$. Therefore, the generalized Brillouin zone can be described by using a parameter $\theta$ as
\begin{equation}
 \beta=\pm\sqrt{\frac{u'-b_z}{u'+b_z}}e^{-i\theta/2}.\label{beta_eq2}
\end{equation}
Finally, we obtain the non-Bloch Hamiltonian
\begin{eqnarray}
 &{}& H(\beta) = \nonumber\\
 &{}&\left(
 \begin{array}{cc}
  c & u'+b_z+\sqrt{\frac{u'+b_z}{u'-b_z}}e^{i\theta/2} \\
  u'-b_z+\sqrt{\frac{u'-b_z}{u'+b_z}}e^{-i\theta/2} & -c
 \end{array}
 \right). \nonumber\\ \label{non-Bloch}
\end{eqnarray}

The non-Bloch Hamiltonian derived above is obtained by transforming a Hermitian Hamiltonian with a regular matrix. Specifically, we consider the following matrix
\begin{eqnarray}
 P = \left(
 \begin{array}{cc}
  \sqrt{u'+b_z} & 0 \\
  0 & \sqrt{u'-b_z}
 \end{array}
 \right), \label{transform_matrix}
\end{eqnarray}
and the non-Bloch Hamiltonian can be described as
\begin{eqnarray}
 P^{-1}H(\beta)P =  \left(
 \begin{array}{cc}
  c & U + e^{i\theta/2} \\
  U + e^{-i\theta/2} & -c
 \end{array}
 \right), \label{non-Bloch2}
\end{eqnarray}
where $U$ represents $U =\sqrt{(u'-b_z)(u'+b_z)}$. The right-hand side is a Hermitian matrix. Since the eigenvalues and band topology are unchanged under this transformation, the appearance of EPs is prohibited in this non-Bloch Hamiltonian. Therefore, the model \eqref{two_EP} shows no stable exceptional surface modes under the open boundary condition in the $z$ direction. We note that the same argument is valid for the cases $b_x=b_z=0$ and $b_y=b_z=0$. Furthermore, even if $b_{x,y,z}$ are nonzero, since the model has the rotational symmetry in the continuum limit, we can expect the absence of exceptional surface modes under the fully open boundary conditions.

\bibliography{reference}

\end{document}